# Red-emitting fluorescent Organic Light emitting Diodes with low sensitivity to self-quenching

S.Forget[(1)*], S.Chenais[(1)], D.Tondelier[(2)], B.Geffroy[(2)], I.Gozhyk[(3)], M.Lebental[(3)], and E.Ishow[(4)]

(1) Laboratoire de Physique des Lasers, Université Paris 13 / CNRS UMR 7538, Villetaneuse, France
(2) Laboratoire de Physique des Interfaces et Couches Minces, Ecole Polytechnique, CNRS UMR 7647, Palaiseau, France
(3) Laboratoire de Photonique Quantique et Moléculaire, ENS Cachan, CNRS UMR 8537, France
(4) Laboratoire de Photophysique et Photochimie Supramoléculaires et Macromoléculaires, ENS Cachan, CNRS UMR 8531, France

*E-mail : sebastien.forget@univ-paris13.fr

*Abstract*

Concentration quenching is a major impediment to efficient organic light-emitting devices. We herein report on Organic Light-Emitting Diodes (OLEDs) based on a fluorescent amorphous red-emitting starbust triarylamine molecule (4-di(4'-tert-butylbiphenyl-4-yl)amino-4'-dicyanovinylbenzene, named FVIN), exhibiting a very small sensitivity to concentration quenching. OLEDs are fabricated with various doping levels of FVIN into $Alq_3$, and show a remarkably stable external quantum efficiency of 1.5% for doping rates ranging from 5% up to 40%, which strongly relaxes the technological constraints on the doping accuracy. An efficiency of 1% is obtained for a pure undoped active region, along with deep red emission (x=0.6; y=0.35 CIE coordinates). A comparison of FVIN with the archetypal DCM dye is presented in an identical multilayer OLED structure.

## 1. Introduction

Organic Light Emitting Diodes (OLEDs) have proven their potential as efficient and low-cost sources for lighting and flat panel displays[1,2]. Their efficiency strongly depends on the properties of the organic materials used for each layer of the OLED architecture, including light-emitting and charge carrier transporting materials. More specifically, the fluorescence efficiency of blue, green and red emitters is a crucial parameter to design high performance

OLEDs for full-color displays and solid-state lighting applications. In this perspective, one of the greatest challenges remains to find red emitters with saturated colors (CIE coordinates x > 0.62 and y < 0.34), high quantum efficiency, and robustness against thermal evaporation allowing a very fine and repeatable control of the device performances[3-6]. Indeed, most of the existing red fluorophores either exhibit strong dipolar structure to form low-energy charge transfer excited states, or possess extensively π-conjugated backbones. This results in strong Van der Waals intermolecular interactions (charge transfer, π−π stacking) causing crystalline aggregates which are often deleterious for optical applications. A related troublesome effect known as "concentration quenching"[7] arises with the increase in molecular density and the appearance of non-radiative complexes. This well-known effect is serious for small-molecule based OLEDs as it becomes highly detrimental as soon as concentration exceeds a few percents. As a consequence, many red-emitting materials exhibit fluorescence quantum yield close to unity in solution, but become either weakly or even not emissive at all in the solid state. Therefore, a universal solution to solve the concentration quenching problem is to use them as dopants in a host material at a very low doping ratio. The archetypal example of such a structure is given by DCM (4-dicyanomethylene-2-methyl-6-(p-dimethylaminostyryl)-4H-pyran) doped into $Alq_3$ (tris(8-hydroxyquinolinate)aluminum(III)). Red emission can here be obtained through direct excitation of the DCM fluorophores or through Förster energy transfer from the excited $Alq_3$ host matrix. The optimal dopant concentration (resulting from the trade-off between efficient energy transfer and reduced concentration quenching) is often very low, commonly not greater than 2%. In practice, the effective doping range requested to keep the OLED efficiency unchanged is very narrow, and limited to a few tenths of percent of the optimal concentration[3,4,8]. From an experimental point of view, doping is by no means a trivial task and controlling the doping level with this level of accuracy is very challenging. This severe drawback strongly limits the repeatability of devices with identical performances in a mass-production process. To our knowledge, systematic studies about the effects of doping level in red-emitting materials have never been deeply explored.

Such investigations stir a tremendous interest for deep-red-emitting materials having a very limited dependence of their properties on the doping rate (when incorporated in a host matrix), or even being able to emit light efficiently in a non-doped configuration (neat films), which would ultimately simplify manufacturing control.

In this paper, we report on a fluorescent red-emitting unsymmetrical starbust triarylamine molecule exhibiting a very small sensitivity to concentration quenching. This molecule[9] was shown to exhibit laser operation under optical pumping in a neat film configuration[10], suggesting high robustness against concentration quenching. We present in this paper a study of the performances (efficiency, spectrum and CIE coordinates) of this material in OLED structures, with various levels of doping in a host system as well as in a non-doped configuration. Comparison with archetypal DCM-based systems presenting a similar multilayer configuration is presented.

## 2. Chemical properties

Several attempts were reported in the literature to obtain red-emitting organic materials with limited sensitivity to quenching effects. Two strategies are commonly adopted consisting either in elaborating molecules presenting a large Stokes shift so as reabsorption effects in concentrated media are minimized, or in introducing sterically crowded substituents or highly branched backbones to avoid the formation of non-radiative π-stacked aggregates. The first approach has been developed in the case of 4-dicyanomethylene-2-methyl-6-[2-(2,3,6,7-tetrahydro-1H,5H-benzo[ij]quinolizin-8-yl)vinyl]-4H-pyran (DCM2), 4-(dicyanomethylene)-2-t-butyl-6(1,1,7,7-tetramethyljulolidyl-9-enyl)-4H-pyran (DCJTB), or 4,4',4''-tris[2-(4-dicyanomethylene-6-t-butyl-4H-pyran-2-yl)-ethylene]triphenylamine (TDCM), while the second one concerns 3-(N-phenyl-N-p-tolylamino)-9-(N-p-styrylphenyl-N-p-tolylamino)perylene ((PPA)(PSA)Pe)[11], anthracen-9-yl-(8-{4-[anthracen-9-yl-(4-ethylphenyl)amino]-phenyl}-benzo[a]-aceant-hrylen-3-yl)-(4-ethylphenyl)amine (ACEN)[12], bis(4-(N-(1-naphthyl)-phenylamino)phenyl)fumaronitrile (NPAFN)[13], 2,3-bis(N, N-1-naphthyl-phenylamino)-N-methylmaleimide (NPAMLI)[14] or benzo[1,2,5]thiadiazole based polymers (BZTA)[15].

Those materials also need to be amorphous since isotropic, homogeneous and grain-boundaries-free systems favor high OLED performances[16].

To this aim, we retained 4-di(4'-tert-butylbiphenyl-4-yl)amino-4'-dicyanovinylbenzene, named FVIN, for its efficient fluorescence emission in the solid state and glass-forming properties. It was synthesized according to a five-step synthetic protocol which was described in earlier study[9]. The presence of a twisted triphenylamino core substituted by terminal bulky *tert*-butyl groups prevents the molecules from interacting with each others and creating deleterious radiationless aggregates. Additionally, this reduces the risks of forming scattering defects known to limit the device optical performances.

Thermal properties were measured by using differential scanning calorimetry (Perkin Elmer Pyris Diamond) in alumina caps under a nitrogen flow at a scan rate of 10 °C.min$^{-1}$ over the temperature range [25 °C - 250 °C]. Thermal analyses actually revealed a glass transition temperature Tg of 86 °C with no further recrystallization. Accordingly thin films obtained under thermal vacuum evaporation from pre-fused powder at a 10 Ås-1 (P = 10-5 mbar) evaporation scan rate showed excellent time-stable amorphous properties after months of storage at room temperature.

The push-pull character of FVIN induced by the presence of the electron-withdrawing dicyanovinylene moiety ensured the formation of a low-energy charge transfer (CT) singlet excited state $S_1$ which absorbs in the visible around 480 nm. $S_1$ relaxes back to the $S_0$ ground state by emitting red light with a maximum at 630 nm (**fig. 1a**) after having evolved toward a more distorted geometry as evidenced from preliminary femtosecond transient absorption spectroscopy measurements. Consequently, the large Stokes shift (5030 cm$^{-1}$) measured for FVIN avoids strong emission quenching in the solid state due to inner filter effects as usually encountered for most of fluorophores, making the emission intensity of FVIN little sensitive to concentration quenching effects. UV-visible absorption spectra were recorded using a Varian Model Cary 5E spectrophotometer. Corrected emission spectra were obtained using a Spex Fluorolog 1681 spectrofluorimeter. Fluorescence quantum yield in cyclohexane solution was determined from a solution of Coumarine 540 A in EtOH ($\Phi_f$ = 0.38) absorbing equally at the excitation wavelength with an absorbance less than 0.1 to avoid reabsorption effects. Measurements of fluorescence quantum yield in the solid state were carried out using a Jobin-Yvon. Inc spectrofluorimeter (Fluoromax 4) equipped with a hollow integrating sphere coated on its internal surface with a diffusely reflecting material (reflecting efficiency: 95% in the spectral range 250-2500 nm).

From theoretical computations conducted in the gas phase (time-dependent density functional theory TDDFT (B3LYP ; 6-31G(d) as a basis set)), the $S_0$-$S_1$ absorption and emission transitions occuring in the visible range involved the highest occupied molecular orbital (HOMO) and lowest unoccupied molecular orbital (LUMO) located at -5.416 eV and -2.483 eV respectively (see **Fig**. **1b**). Their electronic densities were mainly spread on the triphenylamino core and the dicyanovinylene moiety respectively, hence the strong CT character of the corresponding transition ($\lambda_{max}^{calc}$ = 451 nm) with an oscillator strength f calculated to be 0.813. Additionally, a second intense transition ascribed to $S_0 \rightarrow S_2$ was observed in the UV range of the experimental absorption spectrum around 319 nm, and could be accounted for by the HOMO→LUMO+1 transition (f = 0.635; $\lambda_{max}^{calc}$ = 335 nm) with the LUMO+1 mainly localized onto the biphenyl groups. This latter excited state quantitatively deactivated toward the low-energy $S_1$ state since fluorescence quantum yield $\Phi_f$ measurements ($\Phi_f$ = 0.20 in the solid state) showed no wavelength dependence.

The oxidation properties of FVIN were evaluated by cyclic voltammetry in dichloromethane with 0.1 M NBu$_4$ClO$_4$ as a supporting electrolyte at a 100 mVs$^{-1}$ potential scan rate. Cyclic voltammetry was performed using a EG&G PAR 273 potentiostat interfaced to a PC computer. The reference electrode used was an Ag$^+$/Ag electrode filled with 0.01 M AgNO$_3$. This reference electrode was checked versus ferrocene, as recommended by IUPAC (here E°$_{Fc+/Fc}$ = 0.045 V in dichloromethane with 0.1 M tetrabutylammonium perchlorate purchased from Fluka (puriss). Dichloromethane (SDS, 99.9 %) was used as received. All solutions were deaerated by bubbling with argon for a few minutes prior to electrochemical measurements.

The reversible oxidation of the triphenylamino unit into its cation occurs at a half-wave oxidation potential E°$_{ox}$ = 0.57 V vs ferrocene ($\Delta E_p$ = 72 mV), substantially higher than those of unsubstituted triarylamine as logically expected from the electron-withdrawing effects induced by the dicyanovinylene unit located in the para-phenyl position. A second oxidation wave corresponding to the amino dication could be observed at 1.14 V using a higher scan rate (10 Vs$^{-1}$), albeit less reversible ($\Delta E_p$ = 185 mV) than the first oxidation step.

Obviously, through its oxido-reduction properties, the ambipolar compound FVIN closely resembles electron-transporting materials rather than hole-transporting ones[17]. The energetic and electronic descriptions of the HOMO and LUMO will help us to interpret further the experimental OLED features since both molecular orbitals actively participate to the processes of exciton recombination and charge carrier transport.

**3. OLED studies**

**3.1 OLED structure**

The main objective of this study is to show that the FVIN molecule is an efficient fluorophore under electrical driving. Its potential as a red-emitter in organic light-emitting devices was analyzed in a classical multilayer architecture[18-21] realized by vacuum thermal evaporation on ITO coated glass substrates. The isotropic, homogeneous and grain-boundary free properties of the amorphous FVIN material allows easy evaporation in a standard thermal vacuum evaporator. The multilayer structure, depicted in **fig. 2**, successively consisted of a 10-nm

CuPc (copper phthalocyanine) layer used as a hole-injection layer, a 50-nm NPB (N,N′-di(naphthalene-1-yl)-N,N′-diphenyl-benzidine) layer for hole transport, the 20-nm emissive layer (EML) formed either by doping FVIN into $Alq_3$ (devices A1 to A14, see Table 1) or by pure FVIN layer with different thicknesses (devices B1 (5 nm) and B2 (20 nm)), and finally a 50-nm $Alq_3$ layer for electron injection and transport. The device was completed with a LiF (1.2 nm) and aluminum (200 nm) bilayer cathode. The different devices are summarized in table 1. The respective thicknesses of each layer were designed after optical calculations based on a matrix method[21,22] which takes into account the dispersion of the refractive index of each layer, in order to ensure that the EML is located at the maximum of the optical field (for red wavelengths) resulting from interferences due to microcavity effects. For all devices, the $Alq_3$ layer thickness was modified to maintain the total OLED thickness constant when the EML thickness was changed.

The structure of the diode ensures that the exciton recombination zone is located in the vicinity of the NPB/EML layer. With comparable structures, Tang *et al.* reported a 5-nm-wide recombination zone[8], essentially on the electron injection side of the interface. Indeed, electrons encounter a very high energetic barrier at the NPB/EML interface, which prevents them from entering into the NPB layer to form excitons, and also prevents charge transfer excitons from breaking into excitons inside NPB. This is confirmed in our structure by the fact that no blue NPB emission was observed in any of our diodes.

$Alq_3$ was chosen as a host to ensure a good energy transfer from the host to the red-emitting guest. This efficiency is governed by the Förster energy transfer rate constant which is directly proportional to the fluorescence quantum yield of the energy-donating host and the overlap integral between the normalized emission spectrum of the host and the absorption spectrum of the guest. A useful parameter is the Förster critical radius $R_0$ defined as the distance at which the rate of energy transfer to an adjacent acceptor molecule is equal to the fluorescence decay rate of the donor molecule. $R_0$ is given by[23]:

$$R_0^6 = \frac{9000 \ln 10 \, \kappa^2 \, \Phi_d}{128 \, \pi^5 \, N_a \, n^4} J$$

where $\kappa^2$ is the dipole orientation factor (equal to 0.476 in the case of a rigid medium with randomly dispersed acceptors both in space and orientation[24]), $\Phi_d$ is the donor emission quantum yield, n is the refractive index in the area where the transfer occurs, $N_a$ is the Avogadro number and J is the overlap integral between the normalized fluorescence spectrum of the donor and the molar absorption spectrum of the acceptor.

In our case, $R_0$ is calculated to be 30.2 Å, comparable to that obtained for the archetypal Alq$_3$:DCM blend (31.8 Å). An almost complete energy transfer is thus expected at a doping level of a few percents, leading to efficient diodes.

**3.2 OLED measurements**

The glass substrate covered with ITO was cleaned by sonication and prepared by a UV-ozone treatment. The layers were then deposited by sublimation under high vacuum ($10^{-6}$- $10^{-7}$ mbar) at a rate of 0.1-0.2 nm/s in a thermal evaporator. An *in situ* quartz crystal was used to monitor the thickness of the layer depositions with an accuracy of 5 %. The organic materials and the LiF/Al cathode were deposited in a one-step process without breaking the vacuum. The doping rate was controlled by simultaneous co-evaporation of the host and the dopant. After deposition, all the measurements were performed at room temperature and under ambient atmosphere, without any encapsulation. The current–voltage–luminance (I–V–L) characteristics of the devices were measured with a regulated power supply (ACT100 Fontaine) combined with a multimeter (Keithley 4200) and a calibrated silicon photodiode of area 1 cm² (Hamamatsu). The spectral emission was recorded with a SpectraScan PR650 spectrophotometer, with a spectral resolution of 4 nm. The OLED active area is 0.3 cm².

**3.3 Results and discussion**

*3.3.1 Influence of the doping level on color properties*

To be interesting as red-emitting sources for display applications, organic devices should exhibit emission with CIE coordinates verifying x > 0.6 and y < 0.35. It is then interesting to study the conditions leading to deep-red emission. A bathochromic shift of the main emission maximum from 582 nm to 625 nm is observed when the doping concentration of FVIN is increased from 0.5 % to 28 %, as commonly observed with DCM[25], while the spectrum remained unchanged for doping levels above 30%. The evolution of the electroluminescence spectra is plotted in **fig. 3** for various levels of doping, as well as for undoped devices B1 and B2 (EML made of 100% of FVIN).. When a neat film of FVIN was used as an EML (devices B), a small additional shift of the maximum wavelength was also observed (up to 635 nm for device B1), but the main feature in this case relied on the disappearance of the small remaining band around 500 nm. This band is clearly attributed to residual $Alq_3$ emission, which may originate *a priori* either from incomplete energy transfer from $Alq_3$ excitons to FVIN, or from the direct recombination in the pure $Alq_3$ layer of electrons with holes having flowed through the EML. The fact that $Alq_3$ emission intensity remains almost constant instead of decreasing when increasing the doping level tends to discard the first hypothesis, that is incomplete transfer is not the prevailing phenomenon. To check the second hypothesis, we designed a new configuration (device C) incorporating an additional bathocuproine (BCP) hole-blocking layer: for devices C1 and C2, a 10 nm layer of BCP was used between the EML and $Alq_3$ layers to block both holes and excitons. Two diodes were fabricated, with a $Alq_3$:FVIN EML presenting the same doping level as that of devices A2 and A5 for comparison. We observed the suppression of the $Alq_3$ emission band when BCP was introduced (see **fig. 4**). The green emission observed in the $Alq_3$:FVIN blend consequently stems from holes that do not recombine with electrons in the EML, flow through it within $Alq_3$ and then recombine in the neat $Alq_3$ layer. The absence of a green emission shoulder in the devices made of pure FVIN as EML with no BCP suggests that either FVIN molecules do not efficiently transport holes or alternatively that holes accumulate at the FVIN/$Alq_3$ interface because of the 0.4 eV energetic barrier. We plotted in the **fig. 5** the JVL curves (J is the current density in mA/cm², V the voltage and L the luminance) for a low-doped EML (device A1, 0.5% of FVIN in $Alq_3$) and for a pure FVIN EML (device B1, with a thickness of 20 nm for the neat FVIN layer). Even though the slope efficiency of the non-doped device was smaller, the threshold voltage was found to be roughly the same, indicating that the behavior of the diode is still good with a thick pure FVIN layer inserted in the middle of the device.

The CIE coordinates of the doped OLED were affected by the presence of residual $Alq_3$ green emission, hence the resulting emission was not red enough. However, upon addition of BCP a deep red-emission with CIE coordinates of $x = 0.59$ and $y = 0.40$ was eventually obtained (device C2). When non-doped OLEDs were used, the red emission was even more saturated, with $x = 0.613$ and $y = 0.37$. The evolution of the x and y CIE coordinates as a function of the doping level with or without BCP is presented in **fig. 6**.

It is worth noting that no color shift was observed when varying the injected current for the doped or non-doped device (**fig. 7**). This requirement is essential and sometimes poorly filled with doped systems because of a possible shift of the recombination zone location with increasing injected current.

*3.3.2 Influence of the doping level on OLED efficiency*

The evolution of the external quantum efficiency of the investigated OLEDs as a function of the doping level is presented in **fig. 8**. The external quantum efficiency η increased up to a maximum of 2.1% for a doping level of around 2%. This correspond to a power and current efficiency of 1.7 lm/W and 4.8 cd/A respectively. η then slightly decreased but still stayed high (more than 2%) until the doping level reached 4%, where concentration quenching starts playing a significant role. The most striking feature here is that η remains constant (around 1.5%) over a very broad doping range from 5% up to 40%. This may have important practical implications since setting the doping ratio in this range guarantees both reasonable efficiency and insensitivity to doping level fluctuations. After 40%, the decreasing slope is small, since the efficiency is still 1% for non-doped devices. For comparison purposes, two DCM-based OLEDs with $Alq_3$:DCM as the EML were fabricated at two different levels of doping in DCM (2% and 40%). We can clearly see in **fig. 9** that the decrease of the external quantum efficiency is much faster, down to 0.7% for the 40%-doped DCM-based OLED, which is less than half the value obtained for the FVIN-based OLED with an equivalent doping level.

We have shown with this study that FVIN exhibits high potential for electroluminescent devices. First of all, the amorphous structure of the material with relatively low Tg enables a very easy, controllable and reproducible evaporation process.. Moreover, the structure of the molecule leads to reduced sensitivity to quenching effects: it is possible to achieve high efficiency (up to 2.1%, comparable to the well known $Alq_3$:DCM system), with a constant external quantum efficiency of 1.5% obtained over a range of doping ratios between 5% and 40%, which strongly relaxes the technological constraints on the doping accuracy. Finally, FVIN can be used in neat films (without doping) with standard hole transporting materials such as NPB to form an efficient heterostructure.

**5. Conclusion**

We studied in this paper a new fluorescent electroluminescent material emitting in the red in OLED configuration. We showed that the structure of the molecule provides a reduced sensitivity to quenching effects: in OLED experiments, a constant external quantum efficiency of 1.5% was measured for doping rates ranging from 5% up to 40% in an $Alq_3$ matrix, while a maximum efficiency of 2.1% was obtained for a doping rate of 2%. This represents a strong improvement compared to classical DCM blends where the doping rate must be controlled with an accuracy of less than one percent. The evaporation process can thus be simplified to allow for larger fluctuations in the doping rate in the perspective of a mass production process. We also showed that an OLED containing pure FVIN as the emission layer can still be efficient. In this kind of undoped structure, the evaporation process is straightforward as no doping is required at all. Noteworthy a deep-red emission is obtained with CIE coordinates of (0.62; 0.37), which sets FVIN among the rare electroluminescent small-molecules able to efficiently emit in this color area in neat films. The studies conducted here on concentration quenching effects will help the further fabrication of small molecule-based white OLEDs involving doped or stacked multilayers of blue, green and red emitters.

**Acknowledgements**

The authors acknowledge financial support from the ANR (JC/JC call, "BACHELOR" project) and Paris 13 University (BQR credits). We thank A. Beausset for technical assistance, and J.Zyss for fruitful discussions. F.Miomandre is gratefully thanked for the electrochemical measurements in solution.

Figures :

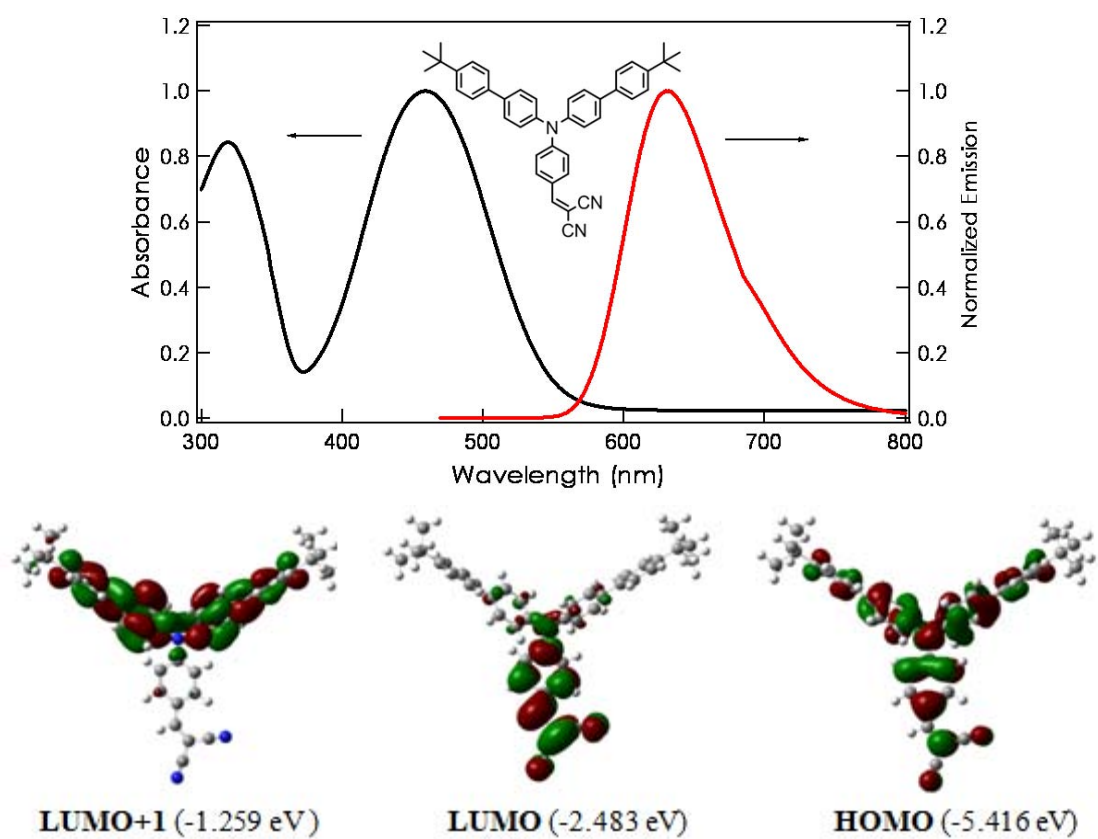

Fig 1 : (a) (from [9]) Normalized absorption and emission spectra of FVIN in thin films (inset : FVIN chemical structure). (b) TDDFT computations (B3LYP, 6-31G(d)) of FVIN in the gas phase. Representation of the electronic density of the HOMO, LUMO and LUMO+1.

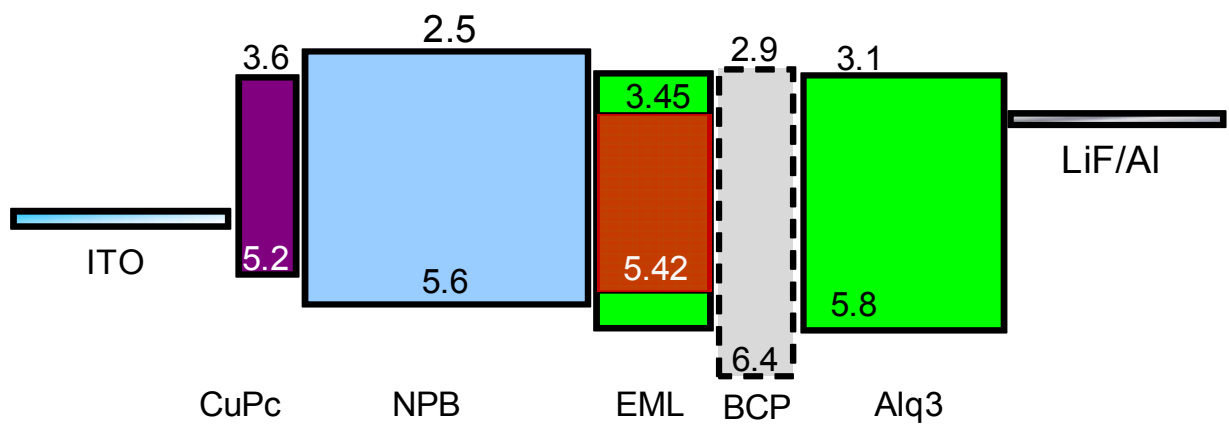

fig 2 : OLED architecture used in the experiments together with HOMO-LUMO levels in eV. For FVIN (EML), the LUMO level is deduced from the HOMO and the optical gap. The other levels are taken from the literature.

The BCP layer in present only in device C. Thicknesses are (10-nm / 50-nm / 20-nm / 50-nm) for the CuPc/NPB/EML/Alq$_3$ layers respectively.

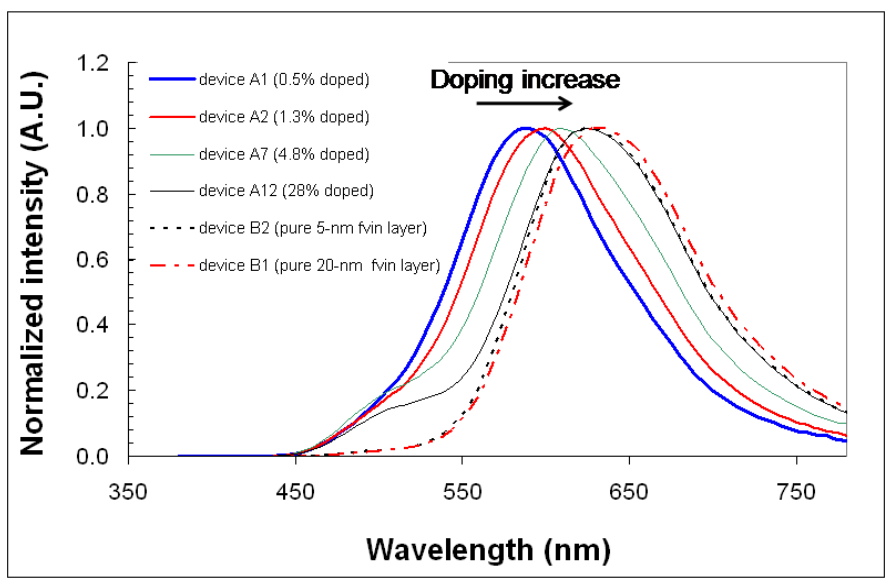

            

fig 3 : evolution of the fluorescence OLED emission spectra in A and B configurations for different doping levels of FVIN in Alq$_3$.

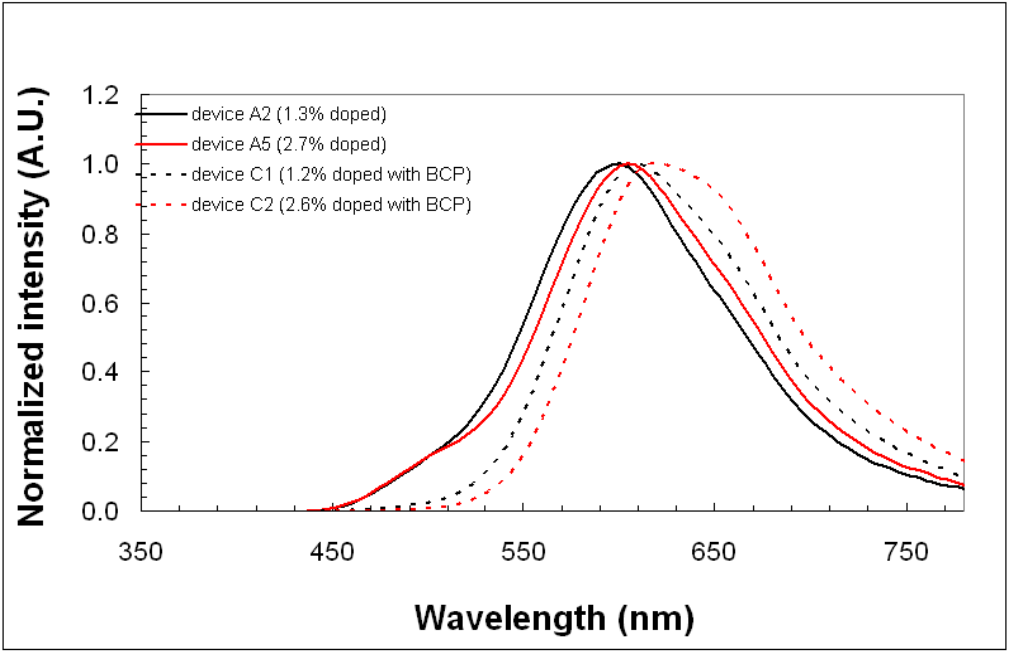

fig 4 : Emission spectra of devices A (without BCP) and C (with BCP) for the same level of doping (here around 1.3 and 2.7 % for devices A2/C1 and A5/C2, respectively).

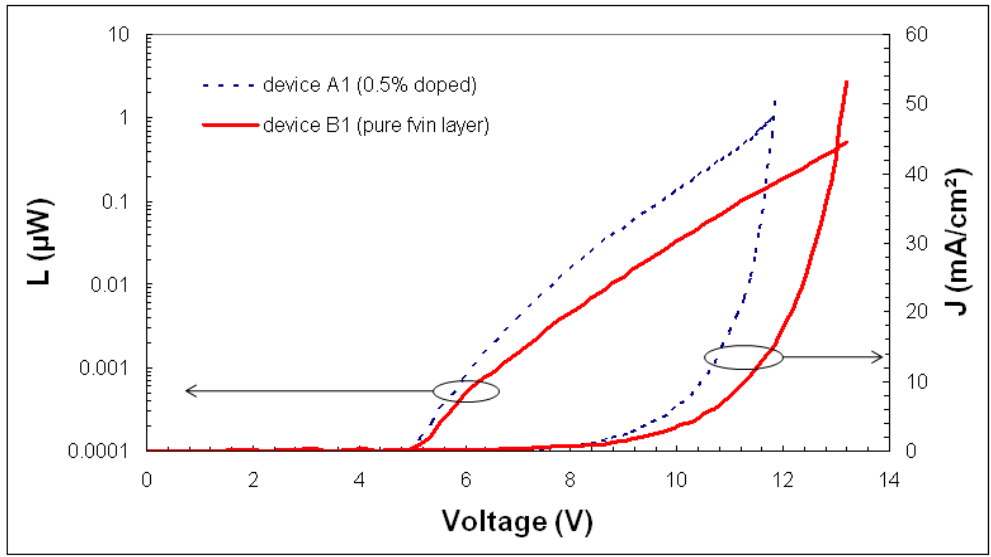

fig 5 : J-V and L-V curves for a low-doped (0.5% of FVIN in $Alq_3$, device A1) and a pure (20 nm neat film, device B1) EML.

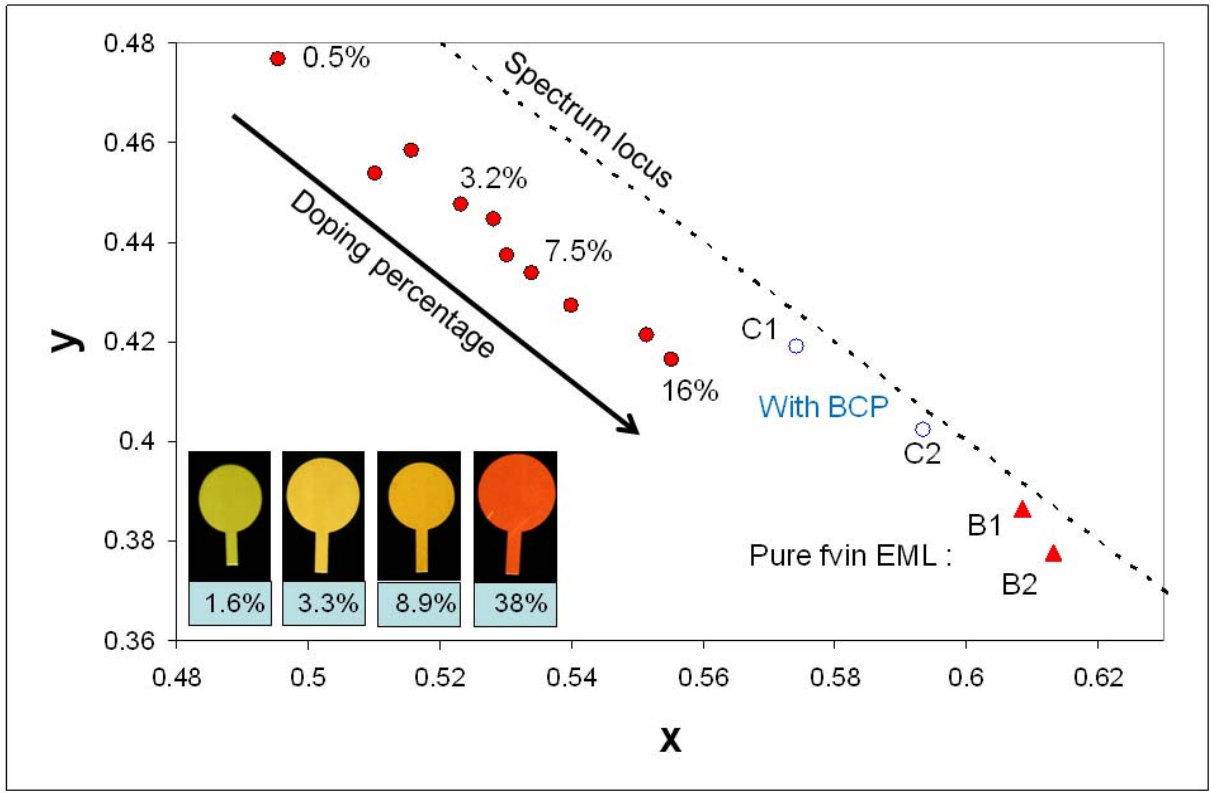

Fig 6 : Evolution of the CIE (x,y) coordinates with the doping level for devices A, B and C. The dots are the doped devices (filled ones without BCP (A), empty ones with BCP (C)) whereas the triangles are the pure-EML devices (B). Inset : photos of four OLEDs with different doping level (color online).

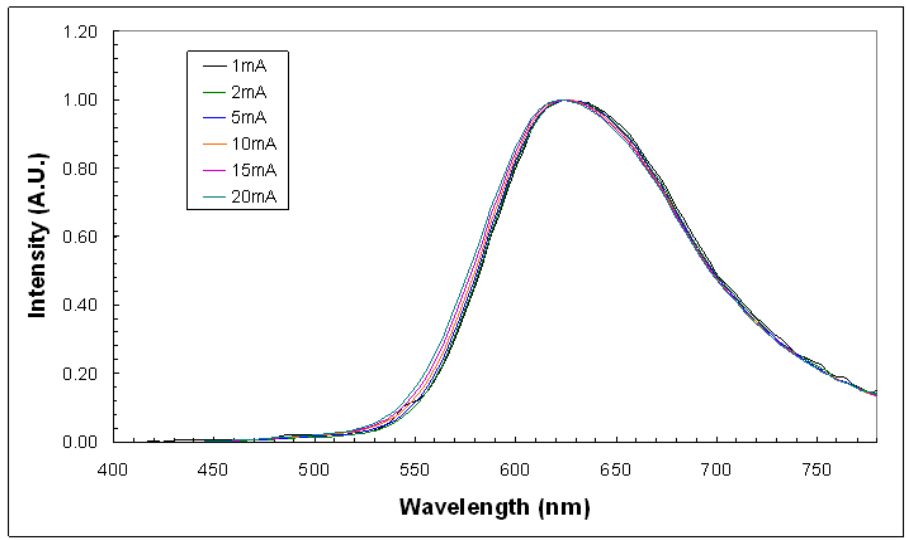

fig 7 : Evolution of the spectrum with injected current for device B1 (20 nm of pure FVIN as EML)

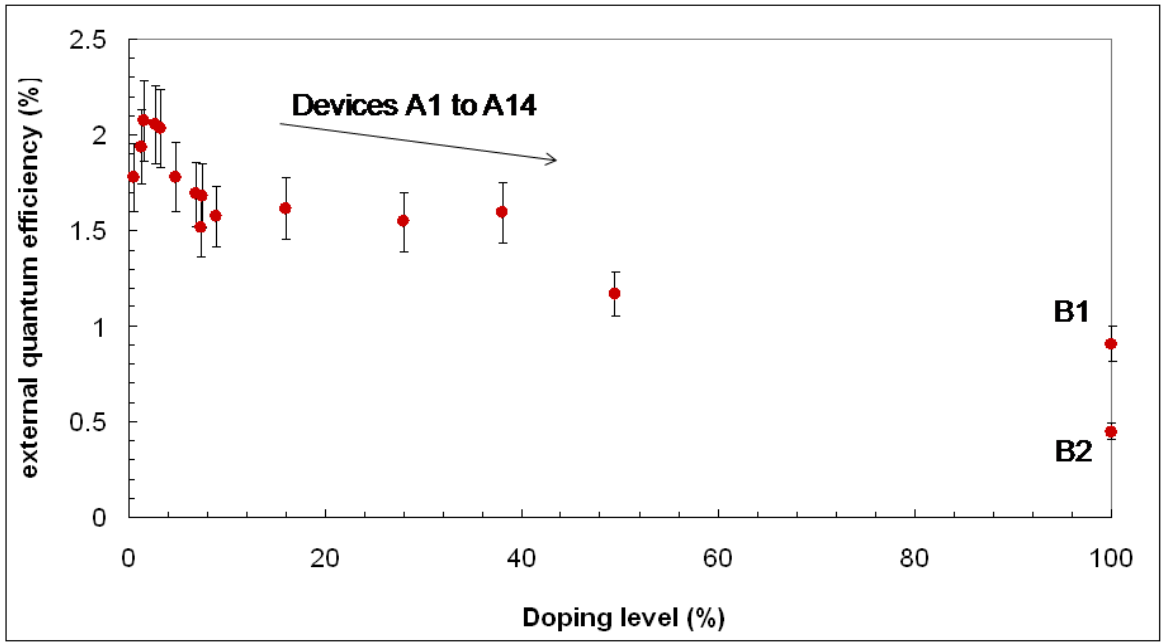

fig 8 : Evolution of the external quantum efficiency with doping level. The devices B1 and B2 used a pure FVIN layer as EML (20 nm and 5 nm wide, respectively)

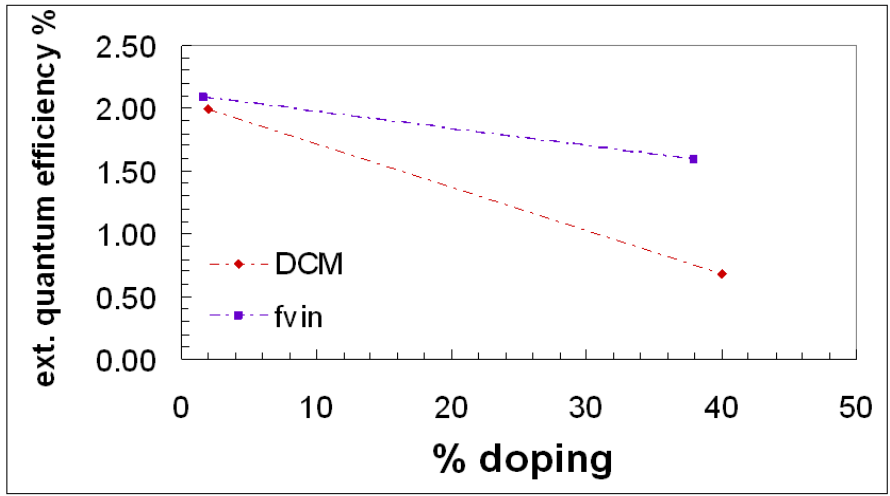

fig 9 : Comparison of external quantum efficiency in FVIN:Alq$_3$ and DCM:Alq3 OLEDs at low and high doping level.

| Name | Doping level (%) of FVIN into Alq$_3$ | Structure |
|------|---------------------------------------|-----------|
|      |                                       |           |
| A1   | 0.5                                   | CuPc/NPB/**Alq$_3$:FVIN**/Alq$_3$ |
| A2   | 1.3                                   | idem      |
| A3   | 2.3                                   | idem      |
| A4   | 2.6                                   | idem      |
| A5   | 2.7                                   | idem      |
| A6   | 3.2                                   | idem      |
| A7   | 4.8                                   | idem      |
| A8   | 6.8                                   | idem      |
| A9   | 7.3                                   | idem      |
| A10  | 7.5                                   | Idem      |
| A11 | 16 | idem |
| A12 | 28 | idem |
| A13 | 32 | idem |
| A14 | 49.5 | idem |
| | | |
| B1 | 100 | CuPc/NPB/**FVIN (20nm)** /Alq$_3$ |
| B2 | 100 | CuPc/NPB/**FVIN (5nm)** /Alq$_3$ |
| | | |
| C1 | 1.2 | CuPc/NPB/Alq$_3$:FVIN/**BCP**/Alq$_3$ |
| C2 | 2.6 | CuPc/NPB/Alq$_3$:FVIN/**BCP**/Alq$_3$ |

Table 1 : list of the different devices